\documentclass[aps,prc,twocolumn,superscriptaddress]{revtex4-2}
\usepackage{graphicx}
\usepackage{makecell}
\usepackage{color}
\usepackage[dvipsnames]{xcolor}
\usepackage[colorlinks=true,urlcolor=Blue,linkcolor=Blue]{hyperref}
\usepackage[all]{hypcap}
\usepackage{amsmath}
\usepackage{ulem}

\begin{document}


\title{$Z=14$ Magicity Revealed by the Mass of the Proton Dripline Nucleus $^{22}$Si}
	
\author{Y.~M.~Xing}
\thanks{These authors contributed equally to this work.}
\affiliation{State Key Laboratory of Heavy Ion Science and Technology, Institute of Modern Physics, Chinese Academy of Sciences, Lanzhou 730000, China}
\affiliation{School of Nuclear Science and Technology, University of Chinese Academy of Sciences, Beijing 100049, China}

\author{Y.~F.~Luo}
\thanks{These authors contributed equally to this work.}
\affiliation{State Key Laboratory of Heavy Ion Science and Technology, Institute of Modern Physics, Chinese Academy of Sciences, Lanzhou 730000, China}
\affiliation{School of Nuclear Science and Technology, University of Chinese Academy of Sciences, Beijing 100049, China}

\author{Y.~H.~Zhang}
\email{Corresponding author: yhzhang@impcas.ac.cn}
\affiliation{State Key Laboratory of Heavy Ion Science and Technology, Institute of Modern Physics, Chinese Academy of Sciences, Lanzhou 730000, China}
\affiliation{School of Nuclear Science and Technology, University of Chinese Academy of Sciences, Beijing 100049, China}	

\author{M.~Wang}
\affiliation{State Key Laboratory of Heavy Ion Science and Technology, Institute of Modern Physics, Chinese Academy of Sciences, Lanzhou 730000, China}
\affiliation{School of Nuclear Science and Technology, University of Chinese Academy of Sciences, Beijing 100049, China}

\author{X.~H.~Zhou}
\affiliation{State Key Laboratory of Heavy Ion Science and Technology, Institute of Modern Physics, Chinese Academy of Sciences, Lanzhou 730000, China}
\affiliation{School of Nuclear Science and Technology, University of Chinese Academy of Sciences, Beijing 100049, China}

\author{J.~G.~Li}
\email{Corresponding author: jianguo\_li@impcas.ac.cn}
\affiliation{State Key Laboratory of Heavy Ion Science and Technology, Institute of Modern Physics, Chinese Academy of Sciences, Lanzhou 730000, China}
\affiliation{School of Nuclear Science and Technology, University of Chinese Academy of Sciences, Beijing 100049, China}

\author{K. H. Li}
\affiliation{State Key Laboratory of Heavy Ion Science and Technology, Institute of Modern Physics, Chinese Academy of Sciences, Lanzhou 730000, China}
\affiliation{Institute of Particle and Nuclear Physics, Henan Normal University, Xinxiang 453007, China}


\author{Q.~Yuan}
\affiliation{State Key Laboratory of Heavy Ion Science and Technology, Institute of Modern Physics, Chinese Academy of Sciences, Lanzhou 730000, China}

\author{Y.~F.~Niu}
\affiliation{School of Nuclear Science and Technology, Lanzhou University, Lanzhou 730000, China}
\affiliation{Frontiers Science Center for Rare Isotope, Lanzhou University, Lanzhou 730000, China}


\author{J.~Y.~Guo}
\affiliation{School of Physics and Optoelectronic Engineering, Anhui University, Hefei 230601, China}


\author{J.~C.~Pei}
\affiliation{State Key Laboratory of Nuclear Physics and Technology, School of Physics, Peking University, Beijing 100871, China}

\author{F.~R.~Xu}
\affiliation{State Key Laboratory of Nuclear Physics and Technology, School of Physics, Peking University, Beijing 100871, China}

\author{G.~de Angelis}
\email{Corresponding author: giacomo.deangelis@lnl.infn.it}
\affiliation{INFN Laboratori Nazionali di Legnaro, viale dell’Università 2 Legnaro, Italy}	
\affiliation{State Key Laboratory of Heavy Ion Science and Technology, Institute of Modern Physics, Chinese Academy of Sciences, Lanzhou 730000, China}

\author{Yu.~A.~Litvinov}
\affiliation{GSI Helmholtzzentrum f{\"u}r Schwerionenforschung, Planckstra{\ss}e 1, 64291 Darmstadt, Germany}

\author{K.~Blaum}
\affiliation{Max-Planck-Institut f\"{u}r Kernphysik, Saupfercheckweg 1, 69117 Heidelberg, Germany}	

\author{I.~Tanihata}
\affiliation{School of Physics, Beihang University, Beijing 100191, China}
\affiliation{Research Center for Nuclear Physics (RCNP), Osaka University, Ibaraki Osaka 567-0047, Japan}

\author{T.~Yamaguchi}
\affiliation{Department of Physics, Saitama University, Saitama 338-8570, Japan}

\author{Y.~Yu}
\affiliation{State Key Laboratory of Heavy Ion Science and Technology, Institute of Modern Physics, Chinese Academy of Sciences, Lanzhou 730000, China}
\affiliation{School of Nuclear Science and Technology, University of Chinese Academy of Sciences, Beijing 100049, China}	

\author{X.~Zhou}
\affiliation{State Key Laboratory of Heavy Ion Science and Technology, Institute of Modern Physics, Chinese Academy of Sciences, Lanzhou 730000, China}
\affiliation{School of Nuclear Science and Technology, University of Chinese Academy of Sciences, Beijing 100049, China}

\author{H.~S.~Xu}
\affiliation{State Key Laboratory of Heavy Ion Science and Technology, Institute of Modern Physics, Chinese Academy of Sciences, Lanzhou 730000, China}
\affiliation{School of Nuclear Science and Technology, University of Chinese Academy of Sciences, Beijing 100049, China}









\author{Z.~Y.~Chen}
\affiliation{State Key Laboratory of Heavy Ion Science and Technology, Institute of Modern Physics, Chinese Academy of Sciences, Lanzhou 730000, China}
\affiliation{School of Nuclear Science and Technology, University of Chinese Academy of Sciences, Beijing 100049, China}	

\author{R.~J.~Chen}
\affiliation{State Key Laboratory of Heavy Ion Science and Technology, Institute of Modern Physics, Chinese Academy of Sciences, Lanzhou 730000, China}
\affiliation{GSI Helmholtzzentrum f{\"u}r Schwerionenforschung, Planckstra{\ss}e 1, 64291 Darmstadt, Germany}

\author{H.~Y.~Deng}
\affiliation{State Key Laboratory of Heavy Ion Science and Technology, Institute of Modern Physics, Chinese Academy of Sciences, Lanzhou 730000, China}
\affiliation{School of Nuclear Science and Technology, University of Chinese Academy of Sciences, Beijing 100049, China}

\author{C.~Y.~Fu}
\affiliation{State Key Laboratory of Heavy Ion Science and Technology, Institute of Modern Physics, Chinese Academy of Sciences, Lanzhou 730000, China}

\author{W.~W.~Ge}
\affiliation{State Key Laboratory of Heavy Ion Science and Technology, Institute of Modern Physics, Chinese Academy of Sciences, Lanzhou 730000, China}

\author{W.~J.~Huang}
\affiliation{Advanced Energy Science and Technology Guangdong Laboratory, Huizhou 516007, China}
\affiliation{State Key Laboratory of Heavy Ion Science and Technology, Institute of Modern Physics, Chinese Academy of Sciences, Lanzhou 730000, China}

\author{H.~Y.~Jiao}
\affiliation{State Key Laboratory of Heavy Ion Science and Technology, Institute of Modern Physics, Chinese Academy of Sciences, Lanzhou 730000, China}
\affiliation{School of Nuclear Science and Technology, University of Chinese Academy of Sciences, Beijing 100049, China}

\author{H.~F.~Li}
\affiliation{State Key Laboratory of Heavy Ion Science and Technology, Institute of Modern Physics, Chinese Academy of Sciences, Lanzhou 730000, China}

\author{T.~Liao}
\affiliation{State Key Laboratory of Heavy Ion Science and Technology, Institute of Modern Physics, Chinese Academy of Sciences, Lanzhou 730000, China}
\affiliation{School of Nuclear Science and Technology, University of Chinese Academy of Sciences, Beijing 100049, China}


\author{J.~Y.~Shi}
\affiliation{State Key Laboratory of Heavy Ion Science and Technology, Institute of Modern Physics, Chinese Academy of Sciences, Lanzhou 730000, China}
\affiliation{School of Nuclear Science and Technology, University of Chinese Academy of Sciences, Beijing 100049, China}

\author{M.~Si}
\affiliation{State Key Laboratory of Heavy Ion Science and Technology, Institute of Modern Physics, Chinese Academy of Sciences, Lanzhou 730000, China}

\author{M.~Z.~Sun}
\affiliation{State Key Laboratory of Heavy Ion Science and Technology, Institute of Modern Physics, Chinese Academy of Sciences, Lanzhou 730000, China}

\author{P.~Shuai}
\affiliation{State Key Laboratory of Heavy Ion Science and Technology, Institute of Modern Physics, Chinese Academy of Sciences, Lanzhou 730000, China}

\author{X.~L.~Tu}
\affiliation{State Key Laboratory of Heavy Ion Science and Technology, Institute of Modern Physics, Chinese Academy of Sciences, Lanzhou 730000, China}

\author{Q.~Wang}
\affiliation{State Key Laboratory of Heavy Ion Science and Technology, Institute of Modern Physics, Chinese Academy of Sciences, Lanzhou 730000, China}

\author{X.~Xu}
\affiliation{State Key Laboratory of Heavy Ion Science and Technology, Institute of Modern Physics, Chinese Academy of Sciences, Lanzhou 730000, China}


\author{X.~L.~Yan}
\affiliation{State Key Laboratory of Heavy Ion Science and Technology, Institute of Modern Physics, Chinese Academy of Sciences, Lanzhou 730000, China}

\author{Y.~J.~Yuan}
\affiliation{State Key Laboratory of Heavy Ion Science and Technology, Institute of Modern Physics, Chinese Academy of Sciences, Lanzhou 730000, China}
\affiliation{School of Nuclear Science and Technology, University of Chinese Academy of Sciences, Beijing 100049, China}


\author{M.~Zhang}
\affiliation{State Key Laboratory of Heavy Ion Science and Technology, Institute of Modern Physics, Chinese Academy of Sciences, Lanzhou 730000, China}

\date{\today}

\begin{abstract}
	
Using the $B\rho$-defined isochronous mass spectrometry technique, we conducted the first mass measurement of the proton dripline nucleus $^{22}$Si. We confirm that $^{22}$Si is bound against particle emission with $S_p/S_{2p}=+1412(114)/+229(54)$ keV, fixing the proton dripline location for the Si element. 
By analyzing the mass differences of the neighboring $sd$-shell nuclei, we find that $^{22}$Si exhibits 
a doubly-magic character similar to its mirror partner $^{22}$O, and that the mirror energy difference of $^{22}$Si-$^{22}$O deviates from the predictions assuming mirror symmetry.
Gamow shell-model calculations reveal that the average occupations of valence protons in $^{22}$Si are nearly identical to those of valence neutrons in $^{22}$O, supporting the $Z=14$ magicity in $^{22}$Si. 
The observed mirror-symmetry breaking is attributed to the extended proton distribution in $^{22}$Si arising from 
a small contribution 
of the unbound $\pi2s_{1/2}$ orbital.
 
\end{abstract}



\maketitle

Determination of the numbers of protons and neutrons that make up magic nuclei has been of paramount importance in understanding the single-particle nuclear shell structure~\cite{PhysRev.75.1969, Haxel}. Characterized by large gaps between single-particle orbitals (shell gaps), single or doubly-magic nuclei play a major role in 
nuclear physics~\cite{YAMAGUCHI2021103882,BLAUM20061,SORLIN2008602,RevModPhys.92.015002} and nuclear astrophysics~\cite{Cowan-2021, Maeda_2022}. The robustness of doubly-magic nuclei makes them not easily excited, and consequently their neighbors can be described in terms of a few extra interacting nucleons, turning a quantum many-body problem into a few-body problem. Recent investigations of nuclei far from stability at radioactive ion-beam facilities have revealed that magic numbers may change in exotic nuclei~{\cite{YAMAGUCHI2021103882,BLAUM20061,SORLIN2008602,RevModPhys.92.015002}}. 
For instance, the conventional magic numbers $N = 8$ and $N = 20$ vanish in the vicinity of $^{12}$Be and $^{32}$Mg~\cite{SORLIN2008602,RevModPhys.92.015002,PhysRevLett.108.142501,MOTOBAYASHI19959,PhysRevLett.99.022503}, while the new magic numbers $N = 14,~16,~32,$ and 34 emerge in $^{22,24}$O~\cite{PhysRevLett.96.012501,THIROLF200016,PhysRevLett.84.5493,PhysRevLett.102.152501,PhysRevLett.109.022501} and $^{36,52,54}$Ca ~\cite{PhysRevLett.131.092501,PhysRevLett.114.202501,RN19234}. Most of these new magic numbers have been identified in the neutron-rich region of the nuclear chart, and thus the question arises how the shell closures or magic numbers evolve at the proton dripline where valence protons are either loosely-bound or unbound.

A fundamental concept in nuclear physics  
	is isospin symmetry, which
has been widely used in theoretical modeling of atomic nuclei~\cite{RN19214}. Based on this concept, nuclei with interchanged protons and neutrons, known as mirror nuclei, exhibit identical characteristics, and magic numbers for protons and neutrons are expected to be the same. For instance, $^{36}$Ca has been found to be a doubly-magic nucleus~\cite{PhysRevLett.131.092501} with a large $N=16$ neutron shell gap compatible with the $Z=16$ proton shell gap in stable $^{36}$S. 
Uniquely, among all the aforementioned exotic doubly-magic neutron-rich nuclei, only $^{22}$O has a proton-rich mirror $^{22}$Si located at the proton dripline~\cite{PhysRevLett.59.33}.

$^{22}$Si has long been a focus of both experimental and theoretical investigations. In the very early times, $^{22}$Si was predicted to be a promising ground-state two-proton emitter~\cite{GOLDANSKY1960482,GOLDANSKY1961648}. However, its discovery at GANIL in 1987 via $^{36}$Ar projectile fragmentation~\cite{PhysRevLett.59.33} , along with the half-life measurement~\cite{PhysRevC.54.572} made it a very marginal candidate for the ground-state $2p$ decay. Recently, the large asymmetry in the Gamow-Teller transitions of $^{22}$Si-$^{22}$O mirror pairs has been reported and suggested as an evidence of proton-halo structure in $^{22}$Al~\cite{PhysRevLett.125.192503}. A subsequent theoretical investigation~\cite{PhysRevC.110.014320} pointed out that the mirror asymmetry between wave functions of $^{22}$Si and $^{22}$O can largely explain the observed asymmetry in the mirror transitions. The mass measurement of $^{22}$Al and the systematic analyses of mirror energy differences (MEDs) do not support the proton-halo structure in the ground state of $^{22}$Al, but in the first excited state\,\cite{PhysRevLett.133.222501}. 
In the case of $^{22}$Si, the proton-halo structure was foreseen in the framework of relativistic mean-field theory~\cite{RN19227,SAXENA2017126}, and state-of-the-art $ab~initio$ valence-space in-medium similarity
renormalization group (VS-IMSRG) calculations showed that $^{22}$Si is a doubly-magic nucleus with both $N=8$ and $Z=14$ shell closures~\cite{LI2023138197}. Up to now, it was uncertain whether $^{22}$Si is particle-bound or -unbound because its experimental mass value was unknown~\cite{SAXENA2017126,PhysRevC.105.034321,KANEKO2017521,PhysRevC.87.014313,AME2020,Kondev_2021}. 

In this Letter, we report the first mass measurement of exotic $^{22}$Si utilizing the newly developed $B\rho$-defined isochronous mass spectrometry ($B\rho$-IMS)\,\cite{PhysRevC.106.L051301,zhangm2023, PhysRevLett.130.192501, PhysRevLett.133.222501}. 
We find that $^{22}$Si
shows a doubly-magic character similar to its mirror partner $^{22}$O.
A neutron skin of 0.43(8) fm has been derived in $^{22}$O\,\cite{PhysRevLett.129.142502}. 
Here we suggest that $^{22}$Si may have an even more extended proton distribution than the neutron distribution in $^{22}$O.


The experiment was conducted at the Heavy Ion Research Facility in Lanzhou (HIRFL)~\cite{XIA200211,ZHAN2010694c}.
$^{36}$Ar$^{15+}$ ions were accelerated to an energy of $E/A\approx445$\,MeV by the main Cooler Storage Ring (CSRm).
The ion beam was fast-extracted and used to bombard a 15-mm thick beryllium target at the entrance of the fragment separator RIBLL2. 
Exotic nuclei were produced via the projectile fragmentation. 
The fragments emerging from the target were fully stripped of bound electrons. 
They were in-flight separated
with RIBLL2, and injected into the 128.8-m long experimental Cooler Storage Ring (CSRe). 

CSRe was tuned to the isochronous mode with $\gamma_t=1.381$, a machine parameter which connects the relative variation of the orbit length, $C$, to the relative change of the magnetic rigidity, $B\rho$, of the circulating ions~\cite{HAUSMANN2000569,FGM}: 
\begin{equation}\label{eq:gammat}
\frac{dC}{C}=\gamma_t^{-2}\cdot\frac{d(B\rho)}{(B\rho)}.
\end{equation}
The momentum acceptance of CSRe was $\pm 0.33 \%$. 
RIBLL2-CSRe was set to a central magnetic rigidity $B\rho=4.8277$\,Tm. 
In this setting, the isochronous condition was optimal for the nuclei with mass-to-charge ratio $m/q\approx1.57$ u (u is the atomic mass unit),
providing the highest mass resolving power and transmission efficiency for $^{22}$Si$^{14+}$.
Every 25\,s, a cocktail beam including the nuclides of interest was injected and stored in CSRe. 
\begin{figure}[htb]
	\includegraphics[angle=0,width=8.5 cm]{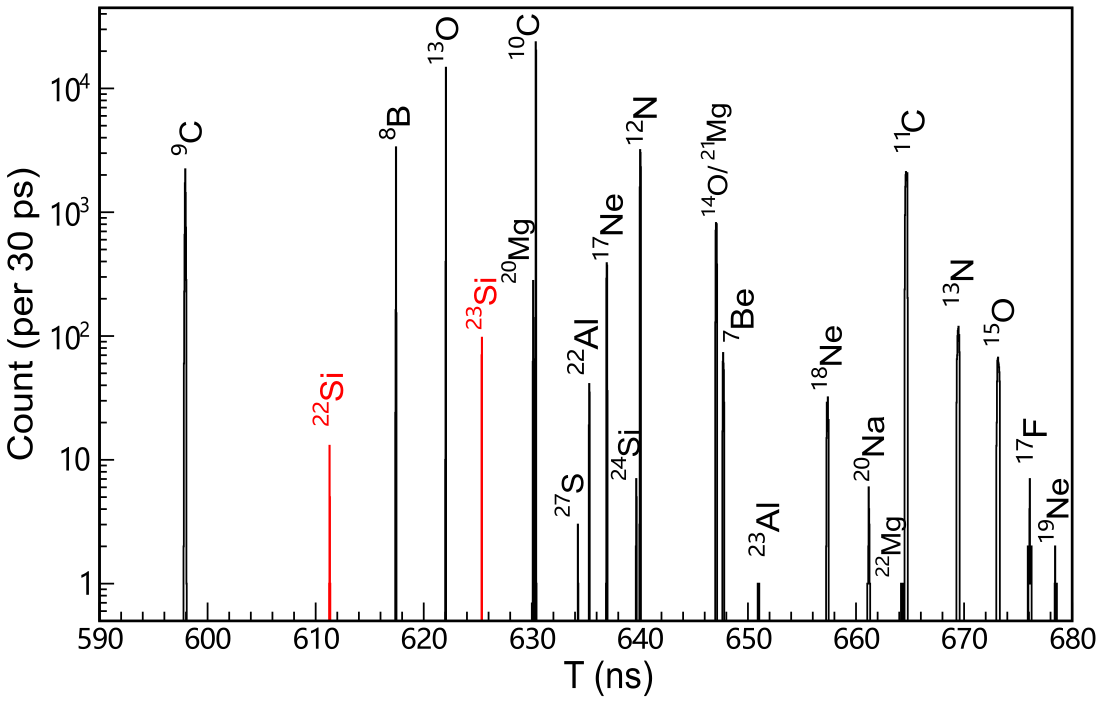}
	\caption{(Colour online) A part of the revolution time spectrum zoomed in the time range of 590 - 680 ns. The red peaks indicate the nuclei of interest.   
		\label{T-spectrum}}
\end{figure}

Two identical time-of-flight (TOF) detectors were installed 18~m apart in one of the two straight sections of CSRe~\cite{ZHANG20141,XLYan2019}.
Each detector consists of a thin carbon foil ($\phi$40~mm, 18~$\rm\mu g/cm^2$) and of a set of micro-channel plates (MCP). 
When an ion passed through the carbon foil, secondary electrons were released from the foil surface and 
guided to the MCP by electric and magnetic fields. 
Fast timing signals from the two MCPs were recorded by a digital oscilloscope at a sampling rate of 50~GHz. 

The measurement duration was 400~$\mu$s after an injection trigger, corresponding to $\sim600$ revolutions of the ions in the ring.
From the timing signals, two sequences of time stamps for each stored ion were extracted\,\cite{XLTu2011N}. Then,
the revolution time, $T$, and velocity, $v$, were determined simultaneously using the procedure described in~\cite{XZhou2021v}. A part of the revolution time spectrum is displayed in Fig.~\ref{T-spectrum}, where particle identification was made following the procedures described in~\cite{XING2019}. 

\begin{table*}[!htbp]
	\caption{Mass excesses (MEs) obtained in this work ${\rm (ME_{IMS})}$. Also listed are the number of events, the literature MEs ($ {\rm ME_{Lit}}$), $\Delta{\rm ME}={\rm ME_{IMS}-ME_{Lit}}$, and the predictions from~\cite{PhysRevC.105.034321} (${\rm ME_{th1}}$), from~\cite{KANEKO2017521}~($ {\rm ME_{INC}}$) using isospin non-conserving forces, and from the improved Garvey-Kelson mass relations~\cite{PhysRevC.87.014313} (${\rm ME_{impGK}}$). 
		The last two columns give one- and two-proton separation energies $S_p$ and $S_{2p}$. The recent result of $S_p(^{21}{\rm Al})=-1.15(10)$ MeV~\cite{PhysRevC.110.L031301} is used to derive the $S_p(^{22}{\rm Si})$ value.}
	\centering
	\footnotesize
	\setlength{\tabcolsep}{4pt}
	\renewcommand{\arraystretch}{1.2}
	\begin{tabular}{lccccccccc}
		\hline
		Atom   &  events  &${\rm ME_{IMS}}$ & ${\rm ME_{Lit}}$ & $\Delta {\rm ME}$ & ${\rm ME_{th1}}$ & ${\rm ME_{INC}}$ & ${\rm ME_{impGK}}$ & $S_p$ & $S_{2p}$  \\
		&          & (keV)  & (keV) & (keV) & (keV) & (keV) & (keV) &(keV) & (keV) \\
		\hline
		$^{22}$Si  &  17  & $31827(54)$    & $33640(500)^{a}$\footnotetext{The ME value is from the extrapolated one in AME20~\cite{AME2020,Kondev_2021}} & $-1813(503) $    & $32128(107)$ & 31886  &  $32051(57)$ & 1412(114) & 229(54) \\
		$^{23}$Si  & 159   & $23365(16)$   & $23537(119)^{b}$\footnotetext{The ME value is from~\cite{PhysRevLett.133.222501} }  &$-172(120)$  & $23240(83) $ &  23411 & 23294(100)  & 2017(16) & 2117(16)  \\
		\hline
	\end{tabular}
	\label{mass values table1}
\end{table*}

\begin{figure}[t]
	\includegraphics[angle=0,width=8.5 cm]{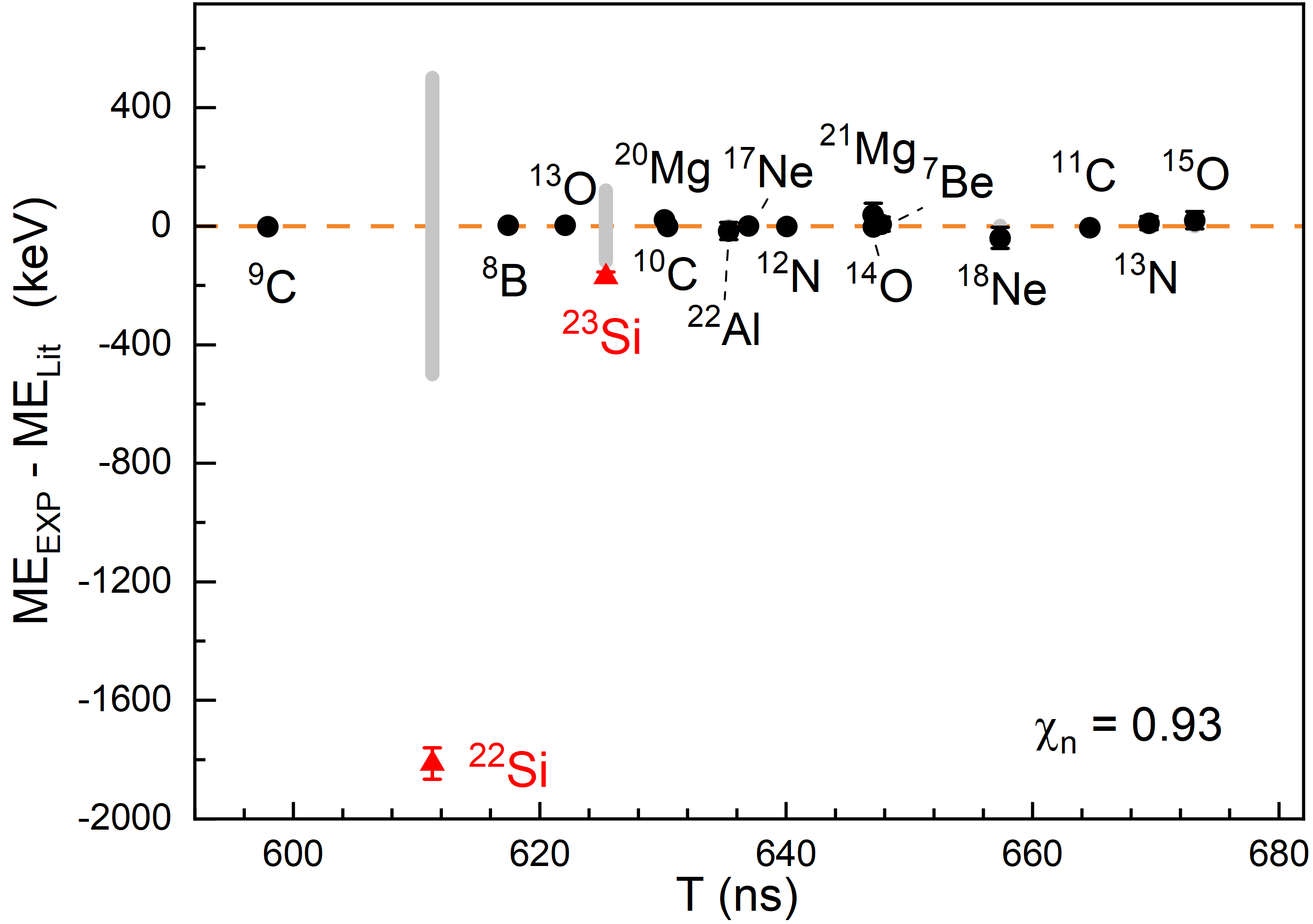}	\caption{(Colour online) 
		Differences of mass excesses (MEs) obtained in this work and those from literature values. The literature MEs for $^{22}$Al and $^{23}$Si are from~\cite{PhysRevLett.132.152501} and~\cite{PhysRevLett.133.222501}, 
		respectively, and the others are from AME20~\cite{AME2020,Kondev_2021}.     
		The new MEs (red triangles) are obtained by utilizing 15 nuclides as calibrants (black circles). Each of these 15 ME values is determined independently using the other 14 nuclides as calibrants and thus demonstrating the reliability of the analysis method. The normalized chi-square $\chi_{n}$ is indicated in the figure. The gray shades represent the $1\sigma$ uncertainties of the literature MEs (not given if smaller than the symbol size).  
		\label{mass_result}}
\end{figure}

Due to the setting on those extremely rarely-produced nuclei, only about 3 ions were stored on average in each injection. Therefore, we had to slightly modify the analysis method described in\,\cite{zhangm2023}.
In the present case, when only one reference ion with well-known mass and the ion of interest were simultaneously stored in an injection, we have determined $T_0$, $v_0 $, $C_0=T_0\cdot v_0$, $(B\rho)_0=(m/q)_0\cdot\gamma_0\cdot v_0$ for the reference ion, and $T_x$, $v_x $, $C_x=T_x\cdot v_x$ for the ion of interest. The slight variation of $\gamma_t$ with orbital length $C$, $\gamma_t(C)$, was derived from the experimental data (see~\cite{ZHANG2022166329} for details). According to Eq.~(\ref{eq:gammat}), we have  
\begin{equation}\label{eq:brhox}
ln\frac{(B\rho)_x}{(B\rho)_0}=\int_{C_0}^{C_x}\gamma_t^2(C)\frac{{\rm d}C}{C}=\sum_{i=1}^n \gamma_t^2(C_i)\frac{\Delta C_s}{C_i}, 
\end{equation}
where $C_i=C_0+i\Delta C_s$, $n=(C_x-C_0)/\Delta C_s$, and $\Delta C_s=0.7$ mm being the numerical integration step. The obtained $(B\rho)_x$ from Eq.~(\ref{eq:brhox}) was used to determine the $m/q$ value of the ion of interest via  
\begin{equation}\label{eq:mass}
\Big(\frac{m}{q}\Big)_x=(B\rho)_x\cdot \sqrt{\frac{1}{v_x^2}-\frac{1}{c^2}}, 
\end{equation}
with $c$ being the speed of light in vacuum. 15 known-mass nuclei with more than 80 counts 
were taken as references.
The obtained mass excesses (MEs) of $^{22,23}$Si and the comparison with several mass predictions~\cite{PhysRevC.87.014313,KANEKO2017521,PhysRevC.105.034321}
are listed in Table~\ref{mass values table1}. 
The comparison to literature values is shown in Fig.~\ref{mass_result}.

The mass of $^{22}$Si is determined for the first time in this work. 
${\rm ME(^{22}Si)} = 31827(54)$\,keV deviates by $\sim1.8$\,MeV from the AME20 extrapolation\,\cite{AME2020,Kondev_2021}, but shows a better agreement with the prediction considering the isospin non-conserving forces~\cite{KANEKO2017521} (see Table~\ref{mass values table1}).
The mass of $^{23}$Si is re-determined with a precision improved by a factor of seven.
The obtained ${\rm ME(^{23}Si)} = 23365(16)$\,keV is consistent within 1.43 standard deviations ($\sigma$) with our previous measurement\,\cite{PhysRevLett.133.222501}. Our new mass value establishes that $^{22}$Si is bound against two-proton emission at a confidence level of $4.2\sigma $ (see Table~\ref{mass values table1}). 
Given that 
$^{21}$Si was not observed in various experiments, it
can be considered to be 
particle unbound~\cite{KANEKO2017521,PFUTZNER2023104050}.
Hence, the location of the proton dripline for Si is herewith fixed.

Since there are no mass data for $^{23}$P and $^{24}$S, 
the conventional mass filters\,\cite{YAMAGUCHI2021103882,BLAUM20061} to probe the magicity of $Z=14$ in $^{22}$Si cannot be applied.
However, as noted in~\cite{PhysRevLett.81.3599}, in a simplified picture,
the differences of nucleon separation energies, $\Delta S_{n,p}$, defined as 
\begin{equation}\label{eq:delta_Sp}
\Delta S_p(Z)=S_p(Z,N)-S_p(Z-1,N), 
\end{equation}
and
\begin{equation}\label{eq:delta_Sn}
\Delta S_n(N)=S_n(Z,N)-S_n(Z,N-1), 
\end{equation}
with even $Z$ and even $N$, can be interpreted as the energy needed to break the outermost nucleon pair. 
This value is sensitive to the level density, and consequently, to the degeneracy of the orbitals occupied by the nucleon pair. 
The persistency of magic numbers can be investigated by examining the systematics
of $\Delta S_{n,p}$. The building up (disappearing) of shell gaps leads to reduced (increased) pairing correlations reflected by the smaller (larger) value of $\Delta S_{n,p}$~~\cite{PhysRevLett.111.162502}.
\begin{figure}[htbp]
	\begin{centering}
		\includegraphics[angle=0,width=8.5 cm]{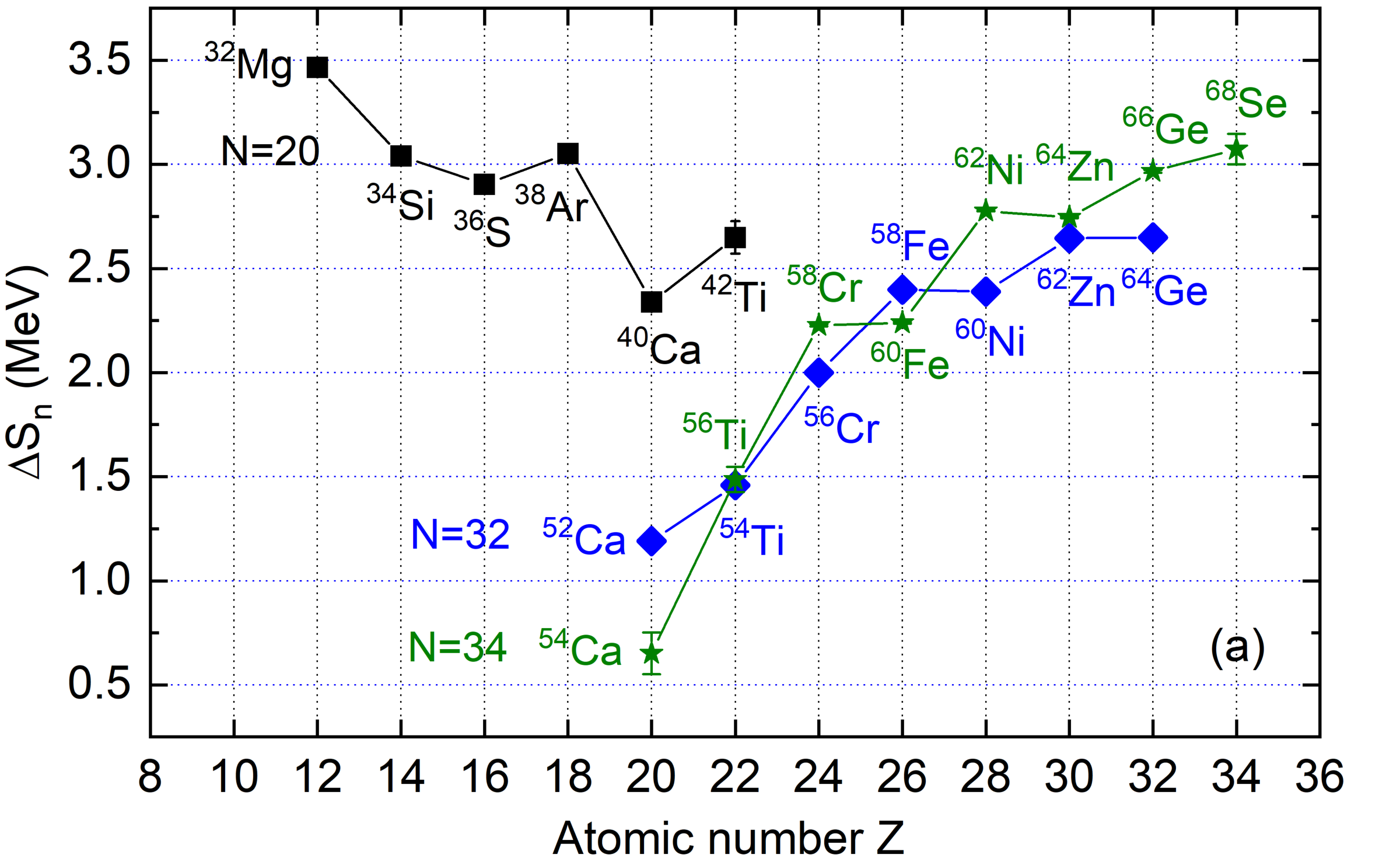}
		\includegraphics[angle=0,width=8.5 cm]{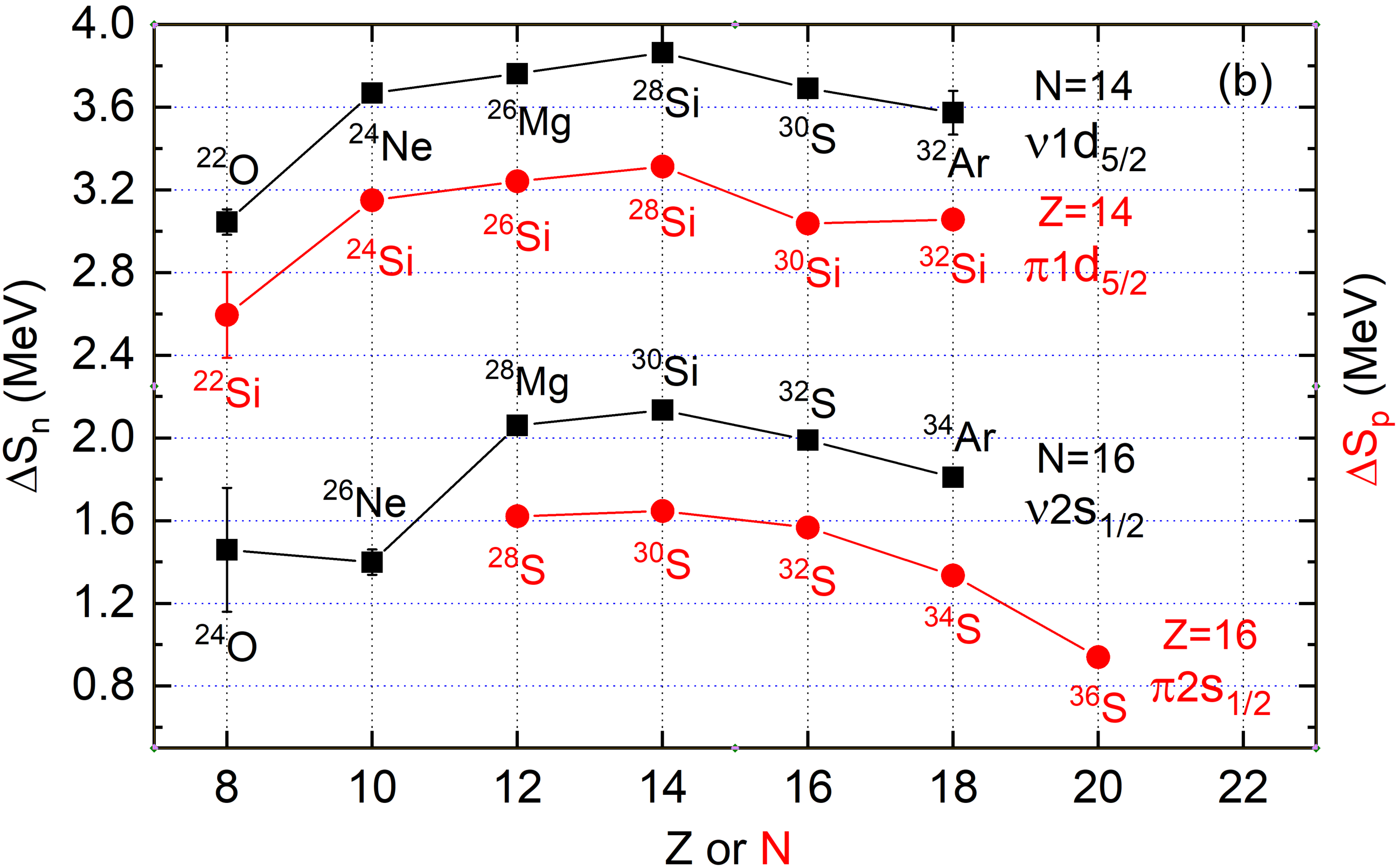}
		\caption{(Colour online) (a) Plots of $\Delta S_n(Z)$
			for $N=20$ (black squares), $N=32$ (blue diamonds), and $N=34$ (green stars) isotones. (b) Same as (a) for $N=14,16$ isotones (black squares), and $\Delta S_p(N)$ for $Z=14,16$ isotopes (red circles). Note the two vertical scales with different colors.
			\label{delta_S1}}
	\end{centering}
\end{figure}

The $\Delta S_n(Z)$ plots deduced from experimental masses are shown in Fig.~\ref{delta_S1}(a) for $N=20$, 32, and 34 isotones, respectively. The disappearance of the conventional $N=20$ magic number in $^{32}$Mg due to deformation is clearly demonstrated by its significant larger $\Delta S_n$ compared to that in $^{40}$Ca. 
For $N=32$ and 34 isotones, the $\Delta S_n$ values become minimal in $^{52}$Ca and $^{54}$Ca due to the effects of tensor forces, which lift significantly the $\nu1f_{5/2}$ orbital and thus make $N=32$ and 34 the
magic numbers~\cite{RN19233}. 
The different numerical $\Delta S_n$ values for $^{40}$Ca, $^{52}$Ca, and $^{54}$Ca are most likely related to the varying occupations of the $\nu1d_{3/2}$, $\nu2p_{3/2}$, and $\nu2p_{1/2}$ single-particle orbitals, respectively~\cite{PhysRevLett.111.162502}.  

To probe the $Z=14$ magicity in $^{22}$Si, 
the plots of $\Delta S_p(N)$ for $Z=14,~16$ isotopes and $\Delta S_n(Z)$ for $N=14,~16$ isotones are presented in Fig.~\ref{delta_S1}(b).
Several features can are evident in this figure. First, the systematic variation of $\Delta S_{p}$ versus $N$ is overall parallel with the variation of $\Delta S_{n}$ versus $Z$, which demonstrates the nearly identical 
orbital occupations in the mirror nuclei~\cite{PhysRevLett.111.162502}.
Second, the energy differences between $\Delta S_n$ and $\Delta S_p$ are around 450 keV primarily due to the repulsive Coulomb interaction of the two paired protons. 
Third, $\Delta S_n$ ($\Delta S_p$) becomes minimum at $^{22}$O and $^{24}$O ($^{22}$Si and $^{36}$S), respectively, along the corresponding isotonic (isotopic) chains, pointing to the shell closures at $N=14,~16$ ($Z=14,~16$). 
The small $\Delta S_n$ value at $^{26}$Ne indicates that the $N=16$ shell closure may also persist in $^{26}$Ne. 
Such systematic trends and the identical behavior of $\Delta S_{n,p}$, especially the minimum values for $^{22}$O and $^{22}$Si, indicate that the latter has a similar, doubly magic, structure. 

The shell evolution at $Z=14$ and $N=14$ was studied via the $ab~initio$ VS-IMSRG calculations, and 
$^{22}$Si was suggested to be a doubly-magic nucleus, similar to its mirror nucleus $^{22}$O, despite bearing a smaller $E_x(2^+_1)$ than that of $^{22}$O~\cite{LI2023138197}. The authors pointed out that the smaller $E_x(2^+_1)$ value in $^{22}$Si is mainly caused by the Thomas-Ehrman shift~\cite{PhysRev.88.1109,PhysRev.81.412} arising from the significant occupation of the $\pi 2s_{1/2}$ orbital. In fact, the low-$l$ orbitals severely affect the structure of loosely bound nuclei. For example, it has been shown that even a slight $s$-orbital occupation can lead to the formation of neutron halo in very weakly bound neutron-rich nuclei~\cite{PhysRevLett.120.052502,PhysRevLett.126.082501}. 

\begin{table}[!htbp]
	\caption{Calculated average orbital occupations of the six valence nucleons outside the $^{16}$O core for $^{22}$Si and $^{22}$O in the ground and $2^+_1$ state. The orbital occupations from $ab~initio$ VS-IMSRG calculations~\cite{LI2023138197} are also given for comparison.}
	\centering
	\footnotesize
	\setlength{\tabcolsep}{4pt}
	\renewcommand{\arraystretch}{1.2}
	\begin{tabular}{lccccccccc}
		\hline
		Nucleus            & $ d_{5/2}$ & $s_{1/2}$ &  $ d_{3/2}$ &  \\
	                       & GSM/$ab~initio $  &  GSM/$ab~initio$ &  GSM/$ab~initio$  &   \\
		\hline
		$^{22}$Si(g.s)     &  5.598/5.528       &   0.195/0.323 &     0.195/0.149 &   \\
    	$^{22}$O(g.s)      &  5.648/5.629 &   0.144/0.226 &     0.199/0.145 &   \\
    	$^{22}$Si($2^+_1$) &  4.757/4.774 &   1.155/1.107 &     0.178/0.119 &   \\
        $^{22}$O($2^+_1$)    &  4.757/4.791 &   1.022/1.084 &    0.212/0.126 &   \\
		\hline
	\end{tabular}
	\label{occupation}
\end{table}

To explore the shell structure and the effects of $s$-orbital occupation in the mirror partners $^{22}$Si-$^{22}$O,
we have performed calculations for $^{22}$Si and $^{22}$O 
using the Gamow shell model (GSM) which utilizes the one-body Berggren basis~\cite{BERGGREN1968265} including bound and resonant states and continuum.
Many-body correlations are incorporated through configuration mixing, while continuum coupling is inherently included at the basis level \cite{PhysRevLett.89.042502,PhysRevLett.89.042501,0954-3899-36-1-013101,Michel_GSM_book,physics3040062}. 
This enables the model to treat both continuum coupling and inter-nucleon correlations simultaneously, making it particularly suitable for describing the exotic properties of drip-line nuclei.
In the calculations,
we take 
$^{16}$O 
as an inert core. The interaction between core and valence nucleons is modeled using a one-body Woods-Saxon potential, while the nucleon-nucleon interaction among the valence nucleons is described using effective field theory forces~\cite{MACHLEIDT20111,RevModPhys.81.1773}.
The constructed GSM Hamiltonian provides a good description for the neutron-rich oxygen isotopes and their mirror nuclei. Also the excitation energy of the $2^+_1$ state in $^{22}$O, $E_x(2^+_1) = 3.2$ MeV~\cite{ENSDF}, is well reproduced.

The GSM calculations reveal that the $\pi 2s_{1/2}$ orbital is unbound in $^{22}$Si while the $\nu 2s_{1/2}$ orbital is deeply bound in $^{22}$O. 
Table~\ref{occupation} presents the calculated average orbital occupations of the six valence nucleons for the ground and $2^+_1$ states in $^{22}$Si and $^{22}$O. These results align closely with the predictions from $ab~initio$ VS-IMSRG calculations~\cite{LI2023138197}. The dominant $1d_{5/2}$ orbital occupations without sizeable mixing with $2s_{1/2}$ and $1d_{3/2}$ indicate that both the $Z=14$ and the $N=14$ shell gaps are enhanced in $^{22}$Si and in $^{22}$O, leading to reduced pairing correlations demonstrated as the minimum values of $\Delta S_{n,p}$, see Fig.~\ref{delta_S1}(b). 
This feature further supports the conclusion that $^{22}$Si and $^{22}$O have similar doubly-magic structure. 

\begin{figure}[t]
	\includegraphics[angle=0,width=8.5 cm]{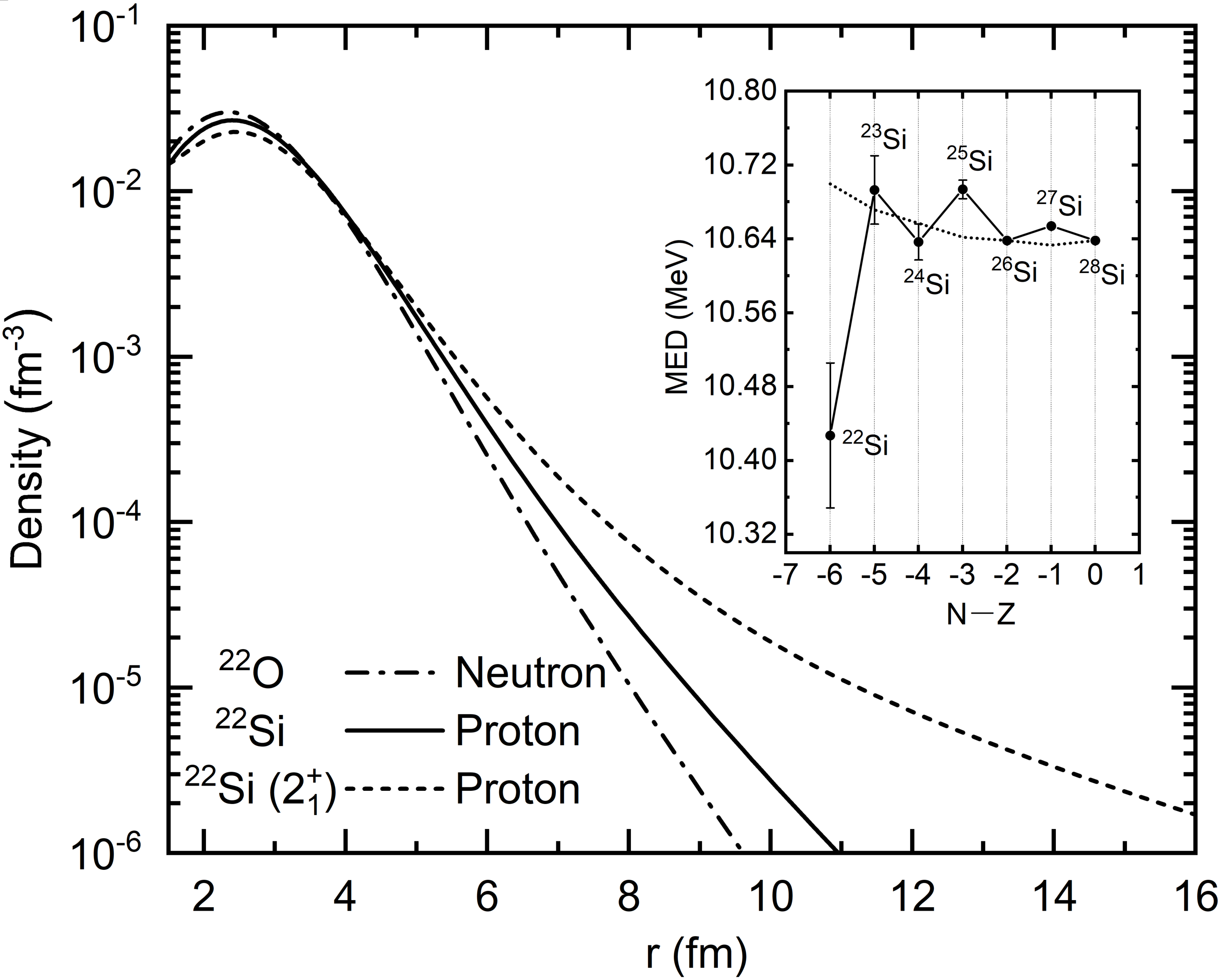}
	\caption{The ground-state proton (neutron) density distribution in $^{22}$Si ($^{22}$O) and the proton density distribution for the $2^+_1$ state in $^{22}$Si obtained from the Gamow shell model calculations. Similar calculations have been done in the case of $^{17}$Ne~\cite{PhysRevLett.101.252502}. The insert shows mirror energy differences (MEDs) versus $N-Z$. The dotted line represents the expected values assuming isospin symmetry~\cite{PhysRevLett.133.222501}.
		\label{density}}
\end{figure}

The valence proton (neutron) density distributions in $^{22}$Si ($^{22}$O) were derived from our GSM calculations and shown in Fig.~\ref{density}. For comparison, the proton density distribution for the first $2^+_1$ excited state in $^{22}$Si is also presented. 
The inset of Fig.~\ref{density} displays the plot of MEDs,
defined~\cite{PhysRevLett.133.222501} as the difference between $S_{2n}$ of the neutron-rich nucleus and $S_{2p}$ of the proton-rich partner. 
The labels in the plot correspond to the proton-rich partners.
Here attention is paid to the sudden drop of MED, i.e., mirror-symmetry breaking in MED, for the $^{22}$Si-$^{22}$O mirror pair, which has been regarded as an indicator of the structure change between the mirror partners~\cite{PhysRevLett.133.222501}.  

From Fig.~\ref{density}, it is evident that the proton density distribution of the $2^+_1$ state in $^{22}$Si has a long tail, which corresponds most probably to the proton-halo structure originating from dominant $\pi 2s_{1/2}$ occupation.   
For the ground state, the $\pi 2s_{1/2}$ occupation is much reduced, and the proton distribution in $^{22}$Si becomes less extended than the $2^+_1$ state, but is  still more extended than that of valence neutrons in 
$^{22}$O (see Fig.~\ref{density}). Such differences in proton and neutron density distributions suggest the presence a proton skin in the ground state of $^{22}$Si, larger than the neutron skin in the ground state of $^{22}$O. Indeed, the neutron-skin thickness in $^{22}$O was determined to be $\Delta R_n=0.43(8)$\,fm~\cite{PhysRevLett.129.142502}, and the proton-skin thickness in $^{22}$Si was predicted to be about 30\% larger than that of the neutron skin in $^{22}$O~\cite{PhysRevLett.130.032501}. 
The slight occupation of the unbound $\pi 2s_{1/2}$ orbital involves the coupling with higher-lying $s_{1/2}$ orbitals from the continuum, driving the nucleus towards a more extended shape. The more extended proton spatial distribution reduces the Coulomb repulsion through the effect of Thomas-Ehrman shift~\cite{PhysRev.88.1109,PhysRev.81.412}
and explains the observed mirror-symmetry breaking of MED in the $^{22}$Si-$^{22}$O mirror nuclei (see the inset of Fig.~\ref{density}). 

In conclusion, the nuclear mass of $^{22}$Si was measured for the first time, and the mass precision of $^{23}$Si was improved by a factor of seven using the newly developed $B\rho$-IMS in CSRe. We confirmed that $^{22}$Si is bound against direct one- and two-proton emissions, thereby establishing the location of proton dripline for Si. The pair-breaking energies, $\Delta S_p$ and $\Delta S_n$, were employed to investigate the shell evolution of mirror nuclei. The systematic trends and the almost identical behavior of $\Delta S_{n,p}$, especially the minimum values for $^{22}$O and $^{22}$Si, indicate that $^{22}$Si has a similar doubly-magic structure as its mirror nucleus $^{22}$O. We performed the Gamow shell model calculations with a full treatment of the continuum effects included. Our calculations revealed that $^{22}$Si not only has the doubly-magic characters but also has a more extended proton distribution in respect to the neutron distribution in 
$^{22}$O; the latter was supported by the observed mirror-symmetry breaking in MED and was attributed to the small $s$-wave component in the wave function 
of $^{22}$Si. 
The coupling to the continuum makes $^{22}$Si a rare example of a doubly-magic nuclear open quantum system\,\cite{Michel_2010}.
Further investigations concerning the measurements of nuclear charge radii and the excitation energy $E_x(2^+_1, {\rm^{22}Si}$), although highly challenging, would be important to reinforce our conclusions.

\begin{acknowledgments}
The authors thank the staff of the accelerator division of IMP for providing stable beam. 
Fruitful discussions with Ivan Mukha are greatly acknowledged.
This work is supported in part by the National Key R\&D Program of China (Grant Nos.~2023YFA1606401 and 2021YFA1601500), the Strategic Priority Research Program of Chinese Academy of Sciences (Grant No.~XDB34000000), the NSFC (Grants No.~12135017, No.~12121005, No.~12475128, No.~12335007, No.~11961141004, No.~12035001, No.~12305126, No.~12205340, No.~12175281, No.~12347106), the Youth Innovation Promotion Association of Chinese Academy of Sciences
(Grants No.~2021419 and No.~2022423), and the Gansu Natural Science Foundation (Grant No.~22JR5RA123, No.~23JRRA614, No.~24JRRA036). G.D.A acknowledges the support by the CAS
President’s International Fellowship Initiative (Grant
No. 2024PVA0006). The theoretical Gamow shell model calculations in this work were done on the Hefei advanced computing center. 
\end{acknowledgments}


\bibliography{Si-22-ref(1)}

\end{document}